\documentstyle{mn}
\title[ Relativistic Cowling Approximation ]
{ Accuracy of the relativistic Cowling approximation \\
 in slowly rotating stars }

\author[  S. Yoshida   and  Y. Kojima   ]
{    Shijun Yoshida$^1$ and  Yasufumi Kojima$^2$ \\
 $^1$Department of Earth Science and Astronomy, 
	 Graduate School of Arts and Sciences, University of Tokyo, \\
         Komaba, Meguro, Tokyo 153, Japan \\
 $^2$Department of Physics, Hiroshima University,  \\
         Higashi-Hiroshima 739, Japan              }

\date{Accepted ?
      Received ?;
      in original form ?}
\pubyear{199?}  

\begin{document}

\maketitle

%--------------------------0

\begin{abstract}
We have calculated the non-radial oscillation 
in slowly rotating relativistic stars with the Cowling approximation.
The frequencies are compared with those based on the complete 
linearized equations of general relativity.
It is found that the results with the approximation 
differ by less than about $20 \%$
for typical relativistic stellar models. 
The approximation is more accurate for higher-order modes 
as in the Newtonian case.
\end{abstract}

\begin{keywords}
oscillation, neutron star
\end{keywords}

\section{ INTRODUCTION  }
%1
    In recent years, our understanding of pulsations of non-rotating 
relativistic stars has been  much improved.
In particular, the role of gravitational wave in the stars 
becomes much clear.
There exists the oscillation mode named as w-mode 
(gravitational {\it w}ave mode), 
associated with the gravitational wave.
The gravitational wave is inherent in the general relativity, so that
the mode becomes evident only for relativistic system.
Except for this new mode, the relativity little affects the modes 
known in the Newtonian pulsation theory. 
The general relativity slightly changes the oscillation frequency
and gives rise to a very slow damping of the mode. 
The emission of gravitational radiation implies that
the oscillation frequency should be complex with a relatively 
tiny imaginary part.
 (See  Andersson, Kojima, and Kokkotas \shortcite{andersson96} 
and  references  therein for the present status of  
the oscillation spectra of non-rotating  relativistic stars.)

%2
The perturbation of the gravity is not so important, 
if the w-mode and the decay of the pulsations 
due to the gravitational radiation do not matter.
This fact hints at the further simplification of the pulsation
problem, that is, neglecting  the perturbation of the 
gravitational field.
This approximation is known as the
Cowling approximation (Cowling \shortcite{cow41})
in the Newtonian stellar pulsation theory,
and gives the same qualitative results and 
reasonable accuracy of the oscillation frequencies.
(See, e.g., Cox \shortcite{cox80}.)
Two different prescriptions, so far, had been proposed
for the relativistic Cowling approximation in non-radial 
pulsations of the non-rotating stars.
One method is that all metric perturbations are neglected, 
i.e., $\delta g_{\mu\nu} = 0$ \cite{mcdermott}. 
The other is that the $\delta g_{rt}$ component of the metric 
perturbations is retained in the pulsation equations \cite{finn}. 
Lindblom and Splinter \shortcite{lindblom90}
examined the accuracy of these two versions of the approximation
for the dipole p-modes,  and concluded 
that the McDermott, Van Horn, and Scholl version is more accurate than 
Finn's version. However,   
the Finn's version is superior in g-mode calculations \cite{finn}. 

%3
In contrast to the oscillations in the non-rotating stars, 
the calculation of oscillation frequencies in rapidly rotating 
relativistic stars seems to be very difficult task. 
The equations become significantly complicated due to 
the relativity and  rotation. See  Priou  \shortcite{priou92} for a set 
of the explicitly written, lengthy equations.
Therefore, the calculation of  normal frequencies  was limited to 
the non-rotating and slowly rotating stars at best \cite{kojima93}.
The possibility to use the Cowling approximation
in rotating relativistic system should be studied.
Ipser and Lindblom \shortcite{ipser92} formulated the Cowling approximation 
of the relativistic system, setting the metric  perturbation to zero.
If the approximation is applied to the stellar pulsation, 
the equation becomes single second order partial differential equation.
The equation is manageable and will be  hopefully solved in the future.
The crucial  point is how good the Cowling approximation 
is to estimate the oscillation frequency in the relativistic regime.
 
%4
In this paper, we will estimate the accuracy of the Cowling approximation
in the frequencies of the non-rotating star and its rotational corrections.
That is, we will compare the frequencies of the Cowling 
approximation with those of the relativistic perturbation theory. 
In section 2, we present the formalism  to calculate normal
frequencies and the rotational shifts by the the Cowling approximation.
In section 3,  the comparison between  both numerical results is given.
Finally, section 4 is devoted to the discussion.
Throughout this paper we will use units of $G = c = 1$.

%
%-------------------------------------------------------------------2%

\section{   METHOD  }

%1
To assess the accuracy of the Cowling approximation,
two distinct calculations are performed. In one calculation,
the relativistic perturbation equations are solved to find 
normal frequencies and the rotational corrections
for the pulsation of stellar models. 
The equations and techniques are described in Kojima (1992,1993).
In the other calculation, we adopt the relativistic Cowling approximation. 
We restrict ourselves to the slowly rotating case,
and write down the equations 
of the pulsation and rotational corrections below.
Normal frequencies are real numbers,
because gravitational perturbations are neglected in 
the Cowling approximation. 
On the other hand, the normal frequencies are 
complex numbers with small imaginary part representing the damping
of the gravitational radiation, when solving 
the exact pulsation equations.
We make a comparison in the oscillation frequency only. 

%-------------------------------------------------------------------2.1%
\subsection{ Equilibrium state }

%1
We assume that the star is slowly rotating with a uniform angular 
velocity $ \Omega $. 
In this paper we consider only the first order effect with 
$ \epsilon = \Omega / \sqrt{ M/R^3 } $,  
where $R$ and $M$ are the radius and the mass of the star, respectively. 
The geometry of a slowly rotating star 
in general relativity is described by
\begin{eqnarray}
ds^2      &=& - e^{\nu} dt^2 + e^{\lambda} dr^2 
     + r^2 (d \theta^2 + \sin^2 \theta \, d \phi^2 ) 
\nonumber \\&&
     - 2 \omega r^2 \sin^2 \theta \, dt d \phi \ , 
\label{metric}\end{eqnarray}
where $\nu$, $\lambda$ and $\omega$ are functions of $r$ only.
Given an equation of state $p = p( \rho ) ,$
the construction of the slowly rotating equilibrium model
is well-known.
(For details see, e.g., Hartle \shortcite{hartle}; 
Chandrasekhar and Miller \shortcite{chandrasekhar74}.)
We will assume a polytropic equation of state,
\begin{equation}
p = k \, \rho^{(1+1/n)} \, ,
\label{eos}\end{equation}
where $n$  and $k$ are constants. 
The function  $\omega  $ is of order $\epsilon , $
and the others, $\nu $, $\lambda $, $p $, 
and $\rho $ are the same as in the non-rotating case.

%---------------------------------------------------------------2.2%
\subsection{ The Cowling approximation }

%1
In this subsection, 
we will explicitly write down the pulsation equation for the slowly 
rotating stars up to the first order of $\epsilon$, using the equations 
derived by Ipser and Lindblom \shortcite{ipser92}. 
We consider only the adiabatic perturbation for the perfect 
fluid, and use the simplest model for the adiabatic index 
$\Gamma$ based on the structure of 
the equilibrium configuration,
\begin{equation}
\Gamma = { \rho + p \over p } \, {d p \over d \rho } \ .
\label{defgamma}\end{equation}
The frequencies of the all $g$-modes  are therefore zero in this case.
The perturbations are given in terms of $e^{-i \sigma t}$, 
radial function $\delta V(r) $, and  
spherical harmonics $Y_{lm} (\theta, \phi)$,
where the azimuthal dependence of the spherical harmonics 
is $e^{i m \phi}$ in our convention.
The oscillation frequency $\sigma $ is the value measured
in the inertial frame. 
Corresponding to equations (16)--(18) of 
Ipser and Lindblom \shortcite{ipser92}, 
the Eulerian changes in the pressure, 
the density, and the four-velocity are 
written up to the first order of $\epsilon$ as
\begin{equation}
\delta p = - \sigma e^{-\nu/2} (\rho + p) \, 
	   \delta V Y_{lm} \, e^{-i \sigma t} \ ,
\label{def1}\end{equation}
\begin{equation}
\delta \rho = - \sigma e^{-\nu/2} \, {(\rho + p)^2 \over \Gamma p}  \, 
	   \delta V Y_{lm} \, e^{-i \sigma t} \ ,
\label{def2}\end{equation}
\begin{equation}
\delta u^t =  - m \, \varpi e^{- \nu} 
          \delta V \, Y_{lm} \, e^{-i \sigma t} \ ,
\label{def3}\end{equation}
\begin{eqnarray}
\delta u^r &=&  i e^{-\lambda} \biggl[ {d \delta V \over dr} + 
	     { \delta V \over \rho + p} \, {d p \over dr} -
\nonumber \\ &&  
             { m  \, e^{\nu} \over \sigma r^2 } \, 
	     {d \over dr} \bigl( r^2 \varpi e^{-\nu} \bigr) \delta V
	     \biggr] \, Y_{lm} \, e^{-i \sigma t} \ ,
\label{def4}\end{eqnarray}
\begin{equation}
\delta u^{\theta} = { i \over r^2} \, \biggl[ 
             \delta V {\partial Y_{lm} \over \partial \theta} - 
             { 2 m \varpi \over \sigma } \delta V
	     \cot \theta \, Y_{lm} \biggr]  \, e^{-i \sigma t} \ ,
\label{def5}\end{equation}
\begin{eqnarray}
\delta u^{\phi} &=&  { - 1 \over r^2 \sin^2 \theta } \Biggl[ 
	     m \, \delta V \, Y_{lm} - 
	     { e^{\nu-\lambda} \over \sigma } \, 
	     {d \over dr } \bigl( r^2 \varpi e^{-\nu} \bigr) \, 
	     \times
\nonumber \\ &&
	     \biggl( {d \delta V \over dr} +  
	     { \delta V \over \rho + p} \, {d p \over dr} \biggr) \, 
	     \sin^2 \theta \, Y_{lm} -
\nonumber \\ &&  
             { 2 \varpi \over \sigma } \,  \delta V \, \sin \theta \, 
	     \cos \theta \, {\partial Y_{lm} \over \partial \theta} 
	     \Biggr] \, e^{-i \sigma t} \ ,
\label{def6}\end{eqnarray}
where 
\begin{equation}
\varpi = \Omega - \omega  .
\end{equation}
The terms with $ \varpi $ are rotational corrections
to the oscillation in the non-rotating star.

Since the background geometry is spherically symmetric, 
the angular part can be separated into each $l, m$ mode.
The basic equation to be solved becomes
second order ordinary differential equation for $\delta V$. 
(See equation (38) of Ipser and Lindblom \shortcite{ipser92}.)
We have
\begin{equation}
\biggl[ {d \over dr} A {d \over dr} + B + \sigma^2 \, C \biggr] \delta V = 
        m \, (2 \sigma \, \Omega \, C - S) \, \delta V \ ,
\label{basiceq1}\end{equation}
where
\begin{equation}
A = (\rho + p) \, r^2 \, e^{(\nu-\lambda)/2} \ , 
\label{basiceq2}\end{equation}
\begin{equation}
B = e^{- \nu/2} \, {d \over dr} 
   \biggl[ r^2 \, {d p \over dr} e^{(2 \nu-\lambda)/2} \biggr] - 
   l(l+1) \, (\rho + p) \, e^{(\nu+\lambda)/2} \ , 
\label{basiceq3}\end{equation}
\begin{equation}
C = { (\rho + p)^2 \over p \Gamma } \, r^2 \, e^{(-\nu+\lambda)/2} \ , 
\label{basiceq4}\end{equation}
\begin{eqnarray}
S  &=& ( \rho + p) \, r^2 \, e^{(\nu+\lambda)/2} \, \Biggl[  
    2 \sigma \varpi e^{ -\nu} + {2 \varpi \over \sigma r^2} - 
\nonumber \\ && 
    {e^{-(3 \nu+\lambda)/2} \over \sigma \, (\rho + p) \, r^2}  \, 
    {d \over dr} \biggl[ (\rho + p) \, e^{(5 \nu-\lambda)/2} \, 
    {d \over dr} \left( \varpi e^{- \nu} r^2 \right) \biggr] \Biggr] \ ,
\nonumber \\
\label{basiceq5}\end{eqnarray}
where $S$ and $\Omega$ are of the order $\epsilon$. 

We will solve the eigen-value problem, equation (\ref{basiceq1}) 
by standard perturbation method, that is,  
expand the potential $\delta V$ and the frequency 
$\sigma$ with respect to the small parameter $\epsilon$ as,
\begin{equation}
\delta V = \delta V_0 + \delta V_1 \, \epsilon + \cdots \ ,
\label{defv}\end{equation}
\begin{equation}
\sigma = \sigma_0 + \sigma_1 \, \epsilon + \cdots \ .
\label{defsigma}\end{equation}
The equation of order $\epsilon^0$  is for the non-rotating case,
\begin{equation}
\biggl[ {d \over dr} A {d \over dr} + B + \sigma_0^2 \, C \biggr] 
       \delta V_0 = 0 \ ,
\label{basiceq11}\end{equation}
and the equation of order $\epsilon $  is 
\begin{eqnarray}
\biggl[ {d \over dr} A {d \over dr} + B + \sigma_0^2 \, C \biggr] 
       \delta V_1 + 2 \sigma_0 \sigma_1 \, C \, \delta V_0 =& & 
\nonumber \\  
        m \sqrt{ { M \over \Omega^2 R^3 }  }  \, 
(2 \sigma_0 \, \Omega \, C - S_0) \, \delta V_0 \ , & &
\label{basiceq12}\end{eqnarray}
where $ S_0 $ is defined by equation (\ref{basiceq5}) 
with $ \sigma = \sigma_0 .$ 
Multiplying $ \delta V_0 ^* $ to equation (\ref{basiceq12}),
and integrating over the star, we have   
the first order rotational correction of 
the normal frequency $\sigma_1$ as
\begin{equation}
\sigma_1 = m \sqrt{ { M \over \Omega^2 R^3 }  }  \, 
    \biggl[ \Omega - 
	  { \int^R_0 S_0 \, {\vert \delta V_0 \vert}^2 \, dr \over 
	   2 \sigma_0 \int^R_0 C \, {\vert \delta V_0 \vert}^2 \, dr } 
             \biggr] \ .
\label{defsigma1}\end{equation}
Similar expression can be found in the Newtonian pulsation equation,
but is written by the  Lagrangian displacement, not by
the Eulerian change of the potential.
Note also that the correction (\ref{defsigma1})
can be determined from the
eigen-function and eigen-value in the non-rotating case, i.e.,
$\delta V_0$ and $\sigma_0$. 
Therefore, it is sufficient to calculate
$\delta V_0$ and $\sigma_0$. 

The frequency in the Cowling approximation should be compared 
with the real part of eigen-frequency calculated by
the exact pulsation equation.
The frequency $\sigma$ of the slowly 
rotating star can be parameterized as
\begin{equation}
\sigma = \sigma_R + \sigma'_R m \epsilon \ , 
\end{equation}
where 
\begin{equation}
\sigma_R = \sigma_0 \ , \ \ \ \sigma'_R = \sigma_1 / m \ .
\end{equation}
The subscript $R$ means the real part of the exact normal frequency.
We will compare results of two calculations
in $\sigma_R $ and $\sigma'_R $.

%---------------------------------------------------------------2.3%
\subsection{ Numerical method }

As mentioned in the last subsection, 
we need eigen-frequency and eigen-function of
the pulsations in the non-rotating stars.
The method to solve equation (\ref{basiceq11}) is well-known and
straightforward. We will briefly summarize the boundary conditions 
and the numerical method. 

%1
Since the basic equation (\ref{basiceq11}) is a second order differential 
equation, two boundary conditions should be given for $\delta V_0$ at the 
center and the surface of the star.
The boundary condition at the center is that
the function $\delta V_0$ should be regular near $ r =0$.
This condition can be written by power series as
\begin{equation}
\delta V_0 = r^l ( v_0 + v_2 \, r^2 + \cdots \ ). 
\label{bounarycondition1}\end{equation}
At the surface, 
the Lagrangian perturbation of pressure should  be vanished,
$ \Delta p = 0 \ . $
This condition can be written explicitly as
\begin{equation}
{d \delta V_0 \over dr } = - \biggl( \sigma_0^2 \, 
e^{\lambda -\nu} (\rho+p) \, { dr  \over dp}  + 
{ 1 \over \rho+p } \, { dp  \over dr}
       \biggr) \, \delta V_0 \ .
\label{bounarycondition2}\end{equation}
%

%2
We can numerically integrate the equation (\ref{basiceq11}) from the 
center with the condition (\ref{bounarycondition1}), and
from the surface with (\ref{bounarycondition2})
to the appropriate interior point.
The eigen-value $\sigma_0$ can be obtained as the consistent frequency 
so as to match functions smoothly at the point.
We also numerically calculate the rotational correction of normal 
frequency  $\sigma_1$,  by $\sigma_0$ and
$\delta V_0$ through the equation (\ref{defsigma1}).

 In order to check  our numerical code for the relativistic Cowling 
approximation, we calculated the fundamental normal mode frequencies 
having harmonic index $l$ in the range $2 \leq l \leq 5$ for the 
non-rotating stellar model used by Ipser and Lindblom \shortcite{ipser92}. 
Our results were in good agreement with their normal frequencies and  
the difference was of the order less than $ 0.05 \% $. 

%---------------------------------------------------------------3

\section{  NUMERICAL RESULTS }

%1
We have calculated normal frequencies for 
a  range of the stellar models with a polytropic  
equation of state (\ref{eos}).
We adapt two different polytropic indices ($ n = 1.0, 1.5$).
The compactness of the star ranges from
$ M/R = 0.05 $ to $ M/R = 0.20 .$
The oscillation frequencies and the rotational corrections are 
calculated for the f-, p$_1$-,  and p$_2$-modes 
having harmonics index $l $ in the range $  2 \leq   l  \leq  5.$ 
The results for typical stellar models are shown in Tables 1--4.
Normal frequencies of the non-rotating stars and the rotational 
corrections are compared.
Since there is not so much difference in the polytropic index,
we hereafter concentrate in  the model with $n =1$.

%2
We show the normal frequencies and rotational corrections 
of the f-modes as a function of $M/R$ in Fig.1.
The eigen-frequencies of the non-rotating star, $\sigma_R$ 
are given in  Fig.1(a), and the rotational corrections $\sigma_R'$
are in  Fig.1(b). The results with the exact relativistic calculation 
are represented by the dashed line, and those with the Cowling 
approximation are by the solid lines.
Figure 1(a) shows that two results of the f-mode approach with 
the increase of $M/R$, that is, the error monotonically decreases.
For example, the relative error for $l=2$ f-mode
in $ M/R=0.05$ model is about $ 30 \% $ in the magnitude,
but it decreases to $ 15 \% $ in $ M/R=0.2$ model.
The relativistic Cowling approximation therefore gives better 
result in the f-mode calculation, as
equilibrium model becomes more relativistic. 
Figure 1(b) shows that both results are always 
in good agreement on the rotational corrections.
The relative error is always less than $ 3 \% $.

%3
Next, we show results of the p$_1$- and p$_2$-modes
in Figs.2 and 3, respectively.
Compared with the exact calculation,
the dependence of $M/R$ is similar in the oscillation frequencies 
and the rotational corrections.
The relative error is almost constant in $\sigma_R$, but increases 
in $\sigma_R'$ with $M/R$. The approximation for these p-modes
becomes worse for the relativistic stars.
However, the discrepancy is always within $10 \%$ error.

%4
From Figs. 1--3, 
it is quite clear that the accuracy of the Cowling approximation 
increases with the spherical harmonic index, $l$.
This feature is well-known in the Newtonian Cowling approximation.
(See, e.g. Cox \shortcite{cox80}.) 
The neglect of the gravitational perturbation
is justified due to the averaging effect of the local fluctuations 
for high-order modes.
The Newtonian Cowling approximation also becomes better
with the increase of the radial node number.
(See, e.g., Robe \shortcite{robe}
for numerical results of the Newtonian Cowling approximation.)
In the relativistic case, the accuracy also increases, 
but the behavior is a little complicated due to the 
additional parameter $M/R$. 

%---------------------------------------------------------------4

\section{  DISCUSSION }

%1
In this paper, we have compared the relativistic Cowling approximation 
with the perturbation calculation of general relativity.  
The accuracy of the Cowling approximation is considerably good,
concerning the normal frequencies in the 
non-rotating star and the first order rotational corrections.
Remarkable fact is that the relative error of the f-mode decreases
with the relativistic factor, $M/R$.
The approximation is also better for larger
radial node number and harmonic index $ l$, 
and greater central condensation,
as in the Newtonian pulsation theory.

%2
One of the important applications of the normal modes
is to determine the critical angular velocity of the rotating stars. 
The maximum angular velocity is limited by the existence of 
the axisymmetric  equilibrium state.
The equilibrium state however suffers secular instability
due to viscosity and/or gravitational radiation reaction (GRR).
The instability  sets in through
$ \sigma = m \Omega $ (viscous instability)
or  $ \sigma = 0 $ (GRR instability),
respectively (e.g., Friedman and Schutz \shortcite{fried78}). 
Form the Newtonian calculation, most significant contributions to 
the GRR instability are the f-modes with $l=-m \sim 4$.
(See, e.g., Yoshida and Eriguchi \shortcite{yoshida95}.) 
The present study shows that the relativistic Cowling 
approximation is able to give a good prediction of the oscillation 
frequency within  $6 \% $ error 
for the f-modes with $l=-m=4, 5$ 
in the typical neutron stars, $M/R \sim 0.2$.
The critical point is however not within the regime of
the linear extrapolation from the non-rotating case,
as suggested in the Newtonian calculation.
That is, the present results, 
$ \sigma_R  $ and  $ \sigma_R '$  are not sufficient, 
and the higher order rotational corrections are necessary.
Therefore, solving normal frequency in
rapidly rotating stars with the Cowling approximation 
would be a good method to determine the critical angular 
velocity.

%\newpage
%-------------------------------------------------------------ack

\section*{  ACKNOWLEDGMENT }
This work was supported in part
by the Grant-in-Aid for Scientific Research Fund of
the Ministry of Education, Science and Culture of Japan 
(08640378).
S.Y. would like to thank Prof. Y. Eriguchi and Prof. T. Futamase for 
their continuous encouragement.

%----------------------------------------------------------------

\newpage
%-------------------------------------------------------------app

%Table for $n=1.0$ and  $M/R=0.1$.
%
\begin{table}
\caption{The normal frequencies of the non-rotating 
stars and the rotational corrections for $n=1.0$ and $M/R=0.1$. 
Subscripts "ex" and "Cow" denote the results with the exact
relativistic perturbation
calculation and the Cowling approximation, respectively.
All entries are given in units of $(M/R^3)^{1/2}$.}
\begin{tabular}{c c c c c c}
\hline \hline
$l$ & mode &  $\sigma_{ex}$ &$\sigma_{Cow}$ &
$\sigma_{ex}'$ &$\sigma_{Cow}'$ \\
\hline
2 & f     & 1.21 & 1.52 &  0.577 &0.589\\    
  & p$_1$ & 3.11 & 3.43 &  0.880 &0.883\\    
  & p$_2$ & 4.83 & 5.09 &  0.925 &0.925\\    
3 & f     & 1.62 & 1.82 &  0.720 &0.723\\
  & p$_1$ & 3.61 & 3.85 &  0.906 &0.908\\    
  & p$_2$ & 5.36 & 5.58 &  0.935 &0.936\\    
4 & f     & 1.93 & 2.07 &  0.789 &0.791\\
  & p$_1$ & 4.03 & 4.22 &  0.923 &0.924\\    
  & p$_2$ & 5.84 & 6.01 &  0.944 &0.944\\    
5 & f     & 2.18 & 2.29 &  0.831 &0.832\\
  & p$_1$ & 4.40 & 4.55 &  0.934 &0.935\\    
  & p$_2$ & 6.27 & 6.42 &  0.950 &0.951\\    
\hline    
\end{tabular}
%\end{table}
%

%
%Table for $n=1.0$ and  $M/R=0.2$.
%\begin{table}
\caption{The same as Table 1, but for $n=1.0$ and  $M/R=0.2$.}
\begin{tabular}{c c c c c c}
\hline \hline
$l$ & mode &  $\sigma_{ex}$ &$\sigma_{Cow}$ &
$\sigma_{ex}'$ &$\sigma_{Cow}'$ \\
\hline
2 & f     & 1.17 & 1.35 &  0.670 &0.680\\    
  & p$_1$ & 2.70 & 2.99 &  0.870 &0.878\\    
  & p$_2$ & 4.12 & 4.38 &  0.898 &0.900\\    
3 & f     & 1.52 & 1.64 &  0.780 &0.782\\
  & p$_1$ & 3.19 & 3.39 &  0.895 &0.900\\    
  & p$_2$ & 4.64 & 4.84 &  0.909 &0.911\\    
4 & f     & 1.78 & 1.87 &  0.832 &0.834\\
  & p$_1$ & 3.59 & 3.74 &  0.912 &0.915\\    
  & p$_2$ & 5.10 & 5.26 &  0.919 &0.920\\    
5 & f     & 2.01 & 2.08 &  0.865 &0.865\\
  & p$_1$ & 3.94 & 4.06 &  0.925 &0.927\\    
  & p$_2$ & 5.52 & 5.65 &  0.927 &0.928\\    
\hline    
\end{tabular}
\end{table}
%

%
%Table for $n=1.5$ and  $M/R=0.1$.
\begin{table}
\caption{The same as Table 1, but for $n=1.5$ and  $M/R=0.1$.}
\begin{tabular}{c c c c c c}
\hline \hline
$l$ & mode &  $\sigma_{ex}$ &$\sigma_{Cow}$ &
$\sigma_{ex}'$ &$\sigma_{Cow}'$ \\
\hline
2 & f     & 1.43 & 1.68 &  0.591 &0.609\\    
  & p$_1$ & 2.87 & 3.18 &  0.851 &0.855\\    
  & p$_2$ & 4.27 & 4.56 &  0.911 &0.911\\    
3 & f     & 1.83 & 1.97 &  0.729 &0.734\\
  & p$_1$ & 3.35 & 3.55 &  0.886 &0.889\\    
  & p$_2$ & 4.76 & 4.98 &  0.926 &0.926\\    
4 & f     & 2.11 & 2.21 &  0.796 &0.798\\
  & p$_1$ & 3.73 & 3.88 &  0.908 &0.910\\    
  & p$_2$ & 5.19 & 5.36 &  0.937 &0.938\\    
5 & f     & 2.34 & 2.41 &  0.835 &0.837\\
  & p$_1$ & 4.05 & 4.16 &  0.923 &0.925\\    
  & p$_2$ & 5.56 & 5.69 &  0.945 &0.946\\    
\hline    
\end{tabular}
%\end{table}
%

%
%Table for $n=1.5$ and  $M/R=0.2$.
%\begin{table}
\caption{The same as Table 1, but for $n=1.5$ and  $M/R=0.2$.}
\begin{tabular}{c c c c c c}
\hline \hline
$l$ & mode &  $\sigma_{ex}$ &$\sigma_{Cow}$ &
$\sigma_{ex}'$ &$\sigma_{Cow}'$ \\
\hline
2 & f     & 1.36 & 1.47 &  0.698 &0.706\\    
  & p$_1$ & 2.46 & 2.71 &  0.855 &0.867\\    
  & p$_2$ & 3.54 & 3.82 &  0.895 &0.900\\    
3 & f     & 1.68 & 1.75 &  0.795 &0.797\\
  & p$_1$ & 2.93 & 3.08 &  0.889 &0.894\\    
  & p$_2$ & 4.05 & 4.24 &  0.909 &0.912\\    
4 & f     & 1.93 & 1.98 &  0.843 &0.844\\
  & p$_1$ & 3.29 & 3.40 &  0.910 &0.912\\    
  & p$_2$ & 4.47 & 4.61 &  0.921 &0.923\\    
5 & f     & 2.13 & 2.17 &  0.872 &0.873\\
  & p$_1$ & 3.60 & 3.68 &  0.924 &0.925\\    
  & p$_2$ & 4.84 & 4.95 &  0.930 &0.932\\    
\hline    
\end{tabular}
\end{table}

\newpage 
$\ $
\newpage 
$\ $
\newpage

\begin{center}{ \bf      Figure caption }\end{center}

\vspace{ 1.cm}
\noindent
Fig.1(a).
%
%Figure
%
Normal frequencies, $\sigma_R$ vs. relativistic factor $M/R$ 
for f-mode and $n=1.0$. Dashed lines and solid lines correspond to 
results with the relativistic perturbation calculation and 
the Cowling approximation, 
respectively. Normal frequencies are normalized by $(M/R^3)^{1/2}$. 
Attached labels $l=2$, $3$, $4$ and $5$ mean the harmonic indices.

%Normal frequencies for 
%non-rotating stars $\sigma_R$ is normalized by $\sqrt{M/R^3}$. 

%
\vspace{ 1.cm}
\noindent
Fig.1(b). 
%
% Figure
%
Rotational corrections, $\sigma_R'$ vs. relativistic factor $M/R$ 
for f-mode and $n=1.0$. Dashed lines and solid lines correspond to 
results with the full calculation and the Cowling approximation, 
respectively. rotational corrections are normalized by $(M/R^3)^{1/2}$. 
Attached labels $l=2$, $3$, $4$ and $5$ mean the harmonic indices.

\vspace{ 1.cm}
\noindent
Fig.2(a). 
%
% Figure
%
The same as Fig. 1(a), but for p$_1$-mode

\vspace{ 1.cm}
\noindent
Fig.2(b). 
%
% Figure
%
The same as Fig. 1(b), but for p$_1$-mode.

\vspace{ 1.cm}
\noindent
Fig.3(a). 
%
% Figure
%
The same as Fig. 2(a), but for p$_2$-mode.

\vspace{ 1.cm}
\noindent
Fig.3(b). 
%
% Figure
%
The same as Fig. 2(b), but for p$_2$-mode.

%-----------------------------------------------------------------%
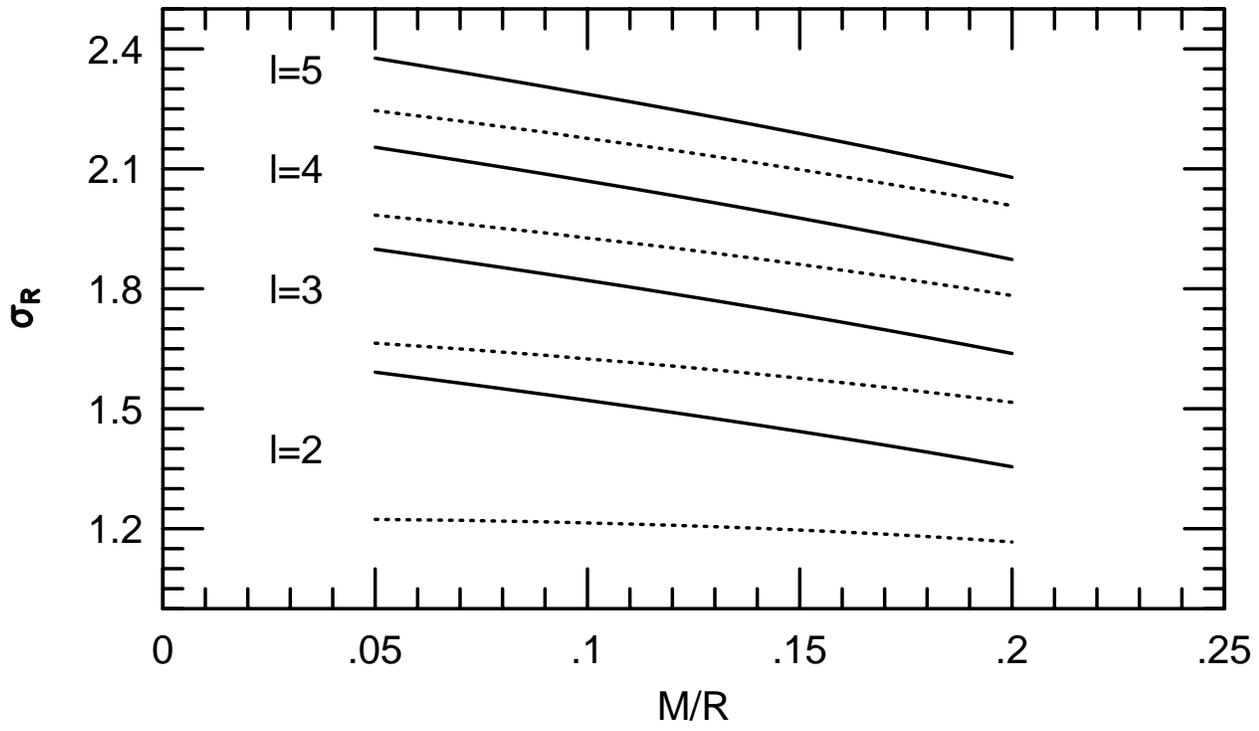
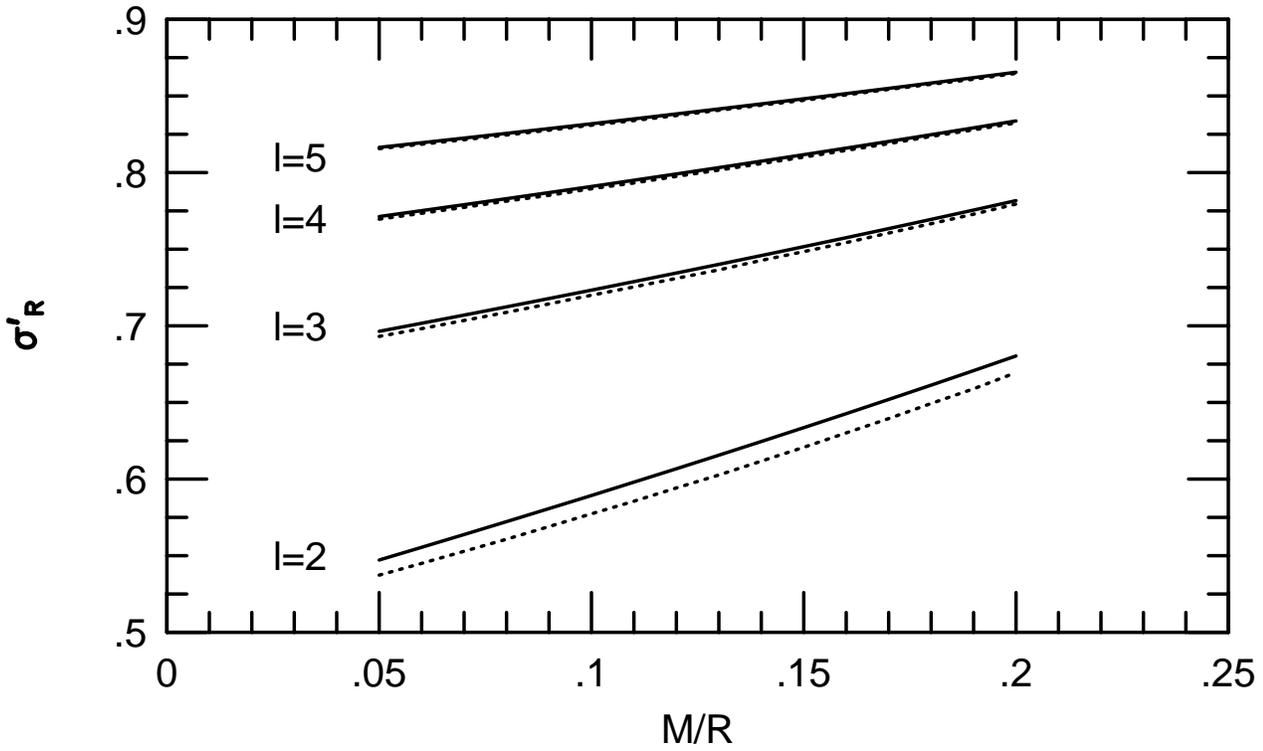
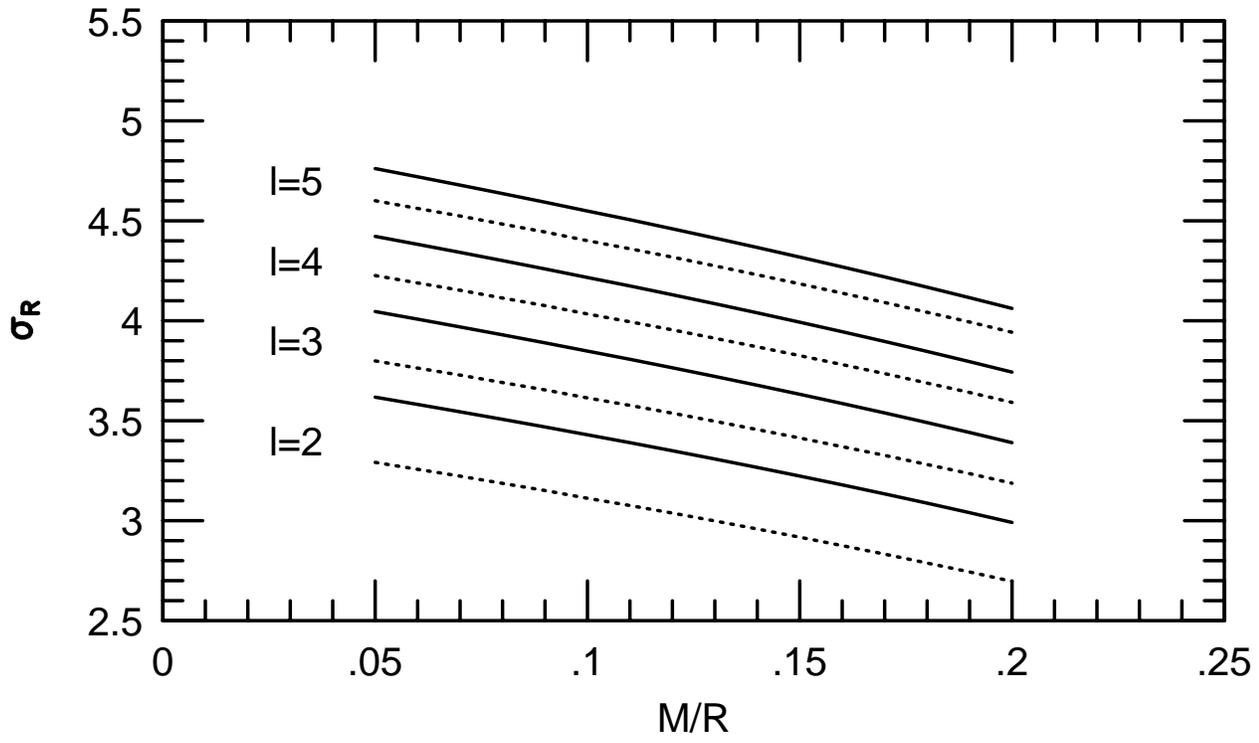
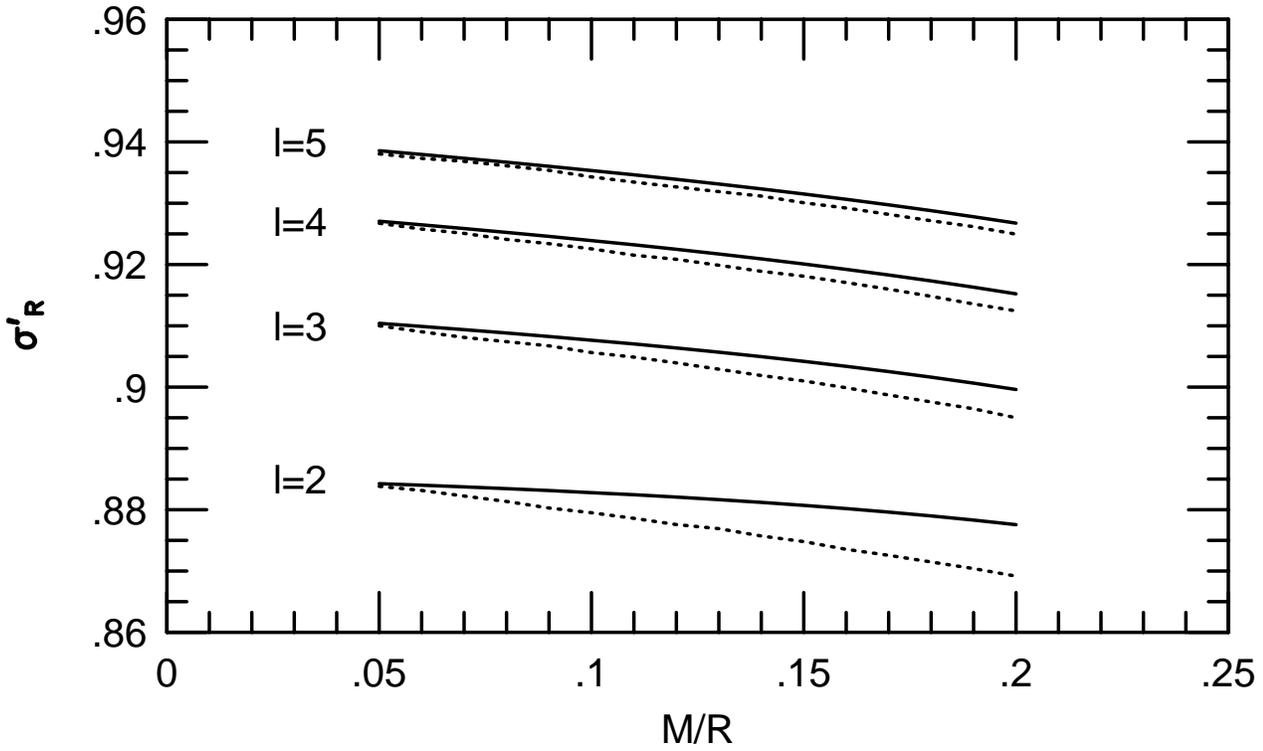
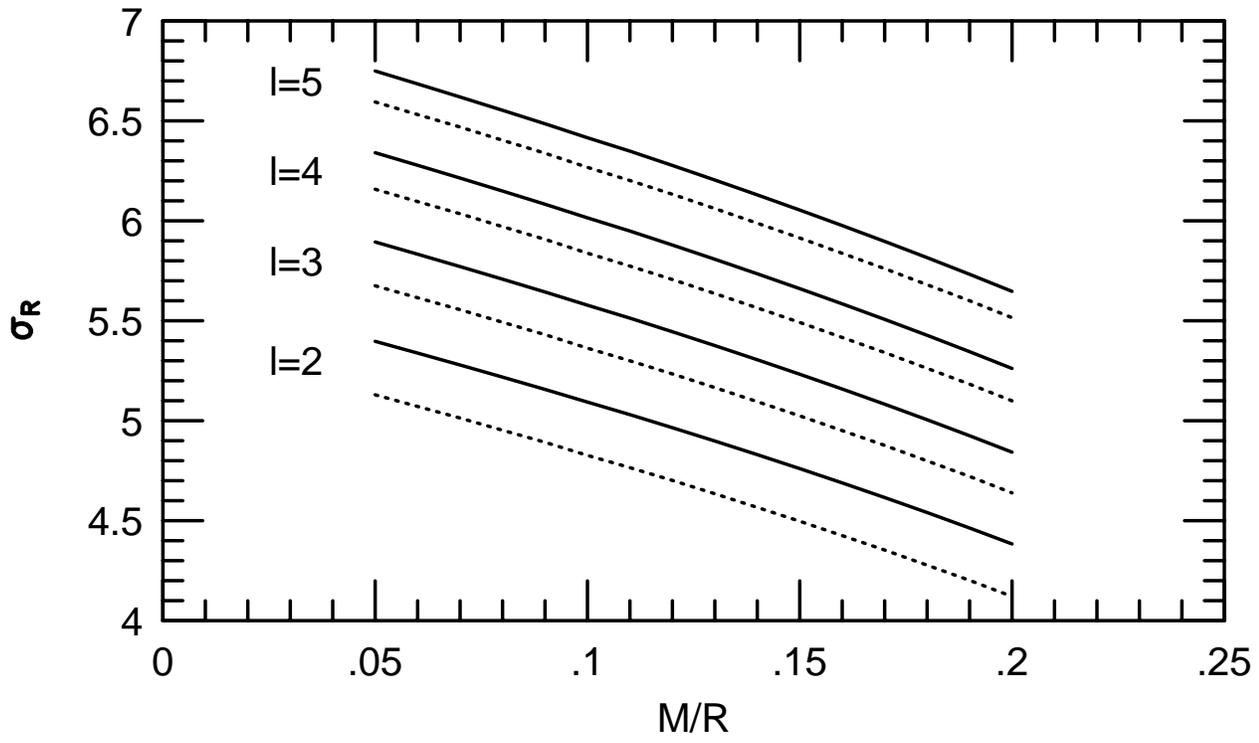
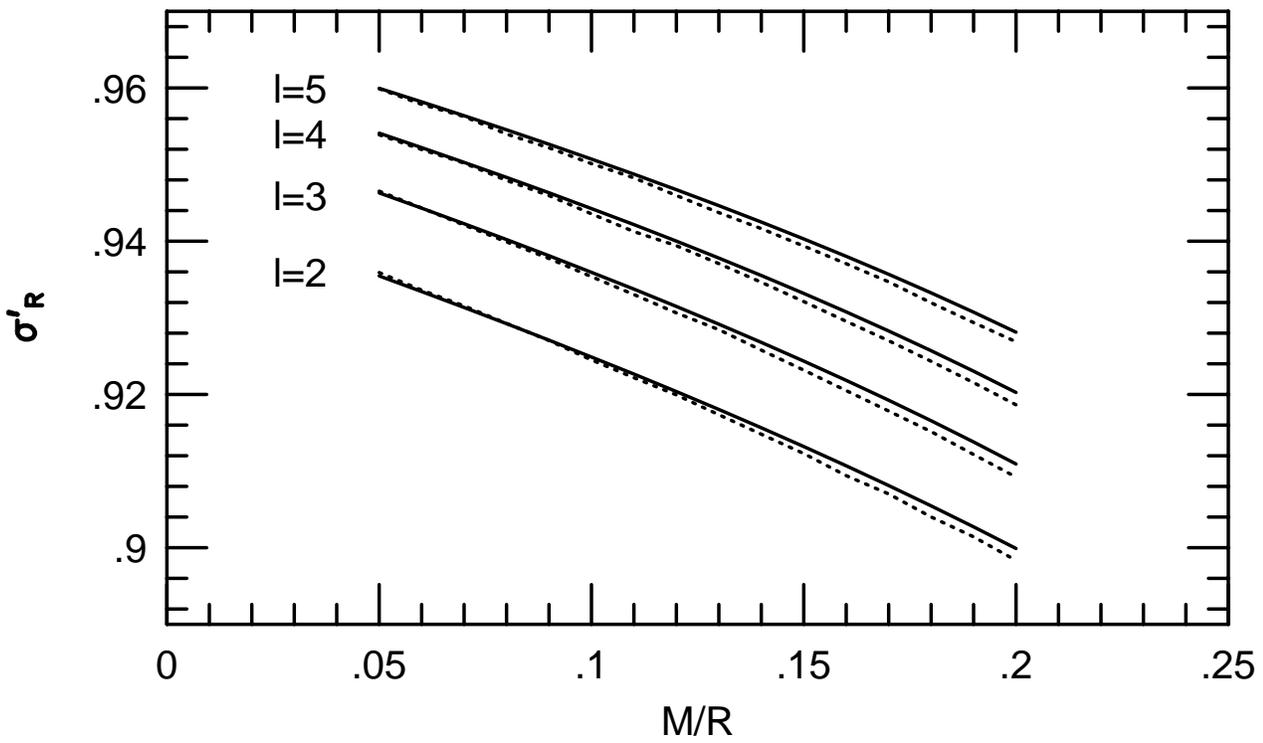
\end{document}